\documentstyle[12pt,epsf,supercite]{article}
\begin{document}
\floatsep -10pt
\textfloatsep 0pt
\intextsep 5pt
\def\textfraction{0.0}
\def\topfraction{1.0}
\def\bottomfraction{1.0}
\setcounter{totalnumber}{20}
\author{ Thomas V. Russo, Richard L. Martin, and\\P. Jeffrey
Hay\\Theoretical Division, MS B268\\Los Alamos National
Laboratory\\Los Alamos, NM 87545\\LA-UR-93-4258}
\title{Density Functional Calculations on First-Row Transition Metals}

\titlepage
\maketitle
\begin{abstract}
The excitation energies and ionization potentials of the atoms in the
first transition series are notoriously difficult to compute
accurately.  Errors in calculated excitation energies can range from
1--4 eV at the Hartree-Fock level, and errors as high as 1.5eV are
encountered for ionization energies.  In the current work we present
and discuss the results of a systematic study of the first transition
series using a spin-restricted Kohn-Sham density-functional method
with the gradient-corrected functionals of Becke and Lee, Yang and
Parr.  Ionization energies are observed to be in good agreement with
experiment, with a mean absolute error of approximately 0.15eV; these
results are comparable to the most accurate calculations to date, the
Quadratic Configuration Interaction (QCISD(T)) calculations of
Raghavachari and Trucks.  Excitation energies are calculated with a
mean error of approximately 0.5eV, compared with $\sim 1\mbox{eV}$ for
the local density approximation and 0.1eV for QCISD(T).  These
gradient-corrected functionals appear to offer an attractive
compromise between accuracy and computational effort.

\end{abstract}


\section{Introduction}
The accurate computation of the excitation energies and ionization
potentials of the first transition metal series has proven to be a
difficult problem for electronic structure theory\cite{reviewdis}.  In
particular, those states which arise from configurations which involve
a doubly occupied $d$ subshell have much larger correlation energies
than those which do not\cite{rlm-hay-xnmet}.  For example, the
difference in correlation energies between the states arising from the
$d^ns^2$ and $d^{n+2}$ occupancies is of the order of 1eV for the
elements Sc--V, where the high-spin $d^{n+2}$ configuration never
involves doubly occupied $d$ orbitals, but increases to $\sim
4\mbox{eV}$ for the elements Mn--Ni, in which the $d^{n+2}$
configuration has two additional filled $d$ sublevels relative to
$d^ns^2$.  The root of this problem lies in the very large Coulomb
interaction and consequent correlation which occurs when two electrons
are required to occupy the same, relatively small, $3d$ orbital.

These errors can, in principle, be eliminated by configuration
interaction (CI) calculations.  Experience has shown
\cite{guse,dunning,martin,bauschlicheretal,lee-freed,rohlfing-martin,%
sunil-jordan,ragha-sunil-jordan,langhoff,salter,rohlfing,bauschlicher,bsp,%
raga-exciting,raga-ions},
however, that quite large one-electron basis sets including angular
momenta at least through $f$ functions are necessary in order to
recover enough of the correlation energy to compute the energy
differences reliably.  Contracted basis sets of dimensions $(10s 7p 4d
3f)$ are necessary to approach the $spdf$ limit.  In addition, the
many-electron basis must be quite large; CI limited to single and
double substitutions yields errors of the order of 0.4--0.7eV for the
$d^ns^2 \rightarrow d^{n+1}s$ excitation energy for the elements to
the right of the series (Mn--Cu), even when a large $spdf$ basis is
used and the results are corrected for relativistic contributions.

The remaining error is associated with higher order excitations, and
therefore several size-extensive methods have been investigated.  The
most standard approach, M\o ller-Plesset perturbation theory, is not
always adequate.  In fact, for those atoms on the right hand side of
the first transition series, the perturbation expansion fails
dramatically and has not converged at fourth
order\cite{rohlfing-martin,raga-exciting,raga-ions} (MP4).  Coupled cluster
techniques, in which certain classes of excitations are summed to
infinite order, give the most reliable results to date.  The Quadratic
Configuration Interaction (QCI) approximation investigated by
Raghavachari and Trucks\cite{raga-exciting,raga-ions} reproduces the
$d^ns^2
\rightarrow d^{n+1}s$ excitation energies with a mean absolute error
of 0.12eV at the QCISD(T) level of theory.  A similar level of
accuracy is obtained for the low-lying ionization potentials.

A number of efforts have also been aimed at examining the performance
of density-functional theory
(DFT)\cite{harris-jones,perdew-zunger,gunnarrsson-jones,harrison,baroni}.
It is somewhat difficult to make a general statement concerning the
level of accuracy achieved by these methods, because the magnitude of
the errors in DFT approaches depend, as expected, on the specific
exchange and correlation functionals used.  A good review of this
area, and the general status of DFT {\em vs.} Hartree-Fock schemes
prior to 1987 is provided by Salahub \cite{reviewdis}.  He points out
that the DFT results also depend on ``whether, and at what stage of
the calculations spherical averaging is invoked.''  While most of the
DFT results appear to be competitive with, or an improvement upon,
Hartree-Fock theory, there are still rather large errors.  Salahub
concludes ``the LSD [local spin density] method generally presents a
reasonable semi-quantitative picture and interprets trends correctly;
however, it yields quantitative errors in relative energies as large
as 1 eV or so.''

Nevertheless, a great deal of useful information has been obtained for
inorganic complexes using the the local density approximation (LDA),
i.e. the Slater exchange functional together with an electron
correlation functional based on the properties of the homogeneous
electron gas (e.g. the Vosko-Wilks-Nusair fit).  Ziegler's recent
review\cite{ziegler} is recommended, as are the articles in the book
edited by Labanowski and Andzelm\cite{labanowski-andzelm}.  The most
significant problem with this type of approach is the tendency of the
LDA to overestimate molecular binding energies, sometimes by as much
as 100\%.  A breakthrough in this regard has been the development of
reliable ``gradient-corrected'' density
functionals\cite{becke-funct,perdew,LYP} (sometimes referred to as
non-local functionals).  In particular, the gradient-corrected
exchange functional of Becke\cite{becke-funct} leads to much improved
bond energies.

Recently a number of studies of gradient-corrected DFT have begun
to
appear\cite{becke96,gjpf,gilljohnpop,dickson-becke,zhu,murray-handy-amos,hma2,%
mlha,andzelm-sosa,johnson,Pople1,becke-grid,scuseria}.
Geometries, vibrational frequencies, dipole moments, and other
properties of small first row molecules are in reasonable agreement
with experiment.  A rule of thumb often mentioned is that these
approaches are comparable in accuracy to MP2 theory.  The
gradient-corrected DFT approaches have one clear advantage, however; the heats
of atomization are remarkable both for their accuracy and for the
fairly modest basis sets required to achieve this accuracy.  It seems
apparent that the density converges much more rapidly with the
one-electron basis than does the correlation energy computed from
Hartree-Fock based approaches.

Given the success of gradient-corrected DFT for molecules composed of
first row atoms, we were curious as to its performance for the
excitation energies and ionization potentials in the first transition
series.  The results of that study are reported in this paper.
Section 2 reviews the methods used to study the Kohn-Sham equations,
including issues regarding the basis sets, integration grids, and the
details of an implementation of spin-restricted Kohn-Sham theory.  The
results are presented and discussed in Section 3, and the conclusions
of this work reiterated in Section 4.

\section{Methodology}
The calculations described here were performed with the MESA suite of
programs\cite{MESA}, using a self-consistent Kohn-Sham (KS)
procedure\cite{KS} with a finite orbital (Cartesian-Gaussian) basis
expansion For the closed shell species we have implemented the
Kohn-Sham equations as described by Pople, Gill and
Johnson\cite{johnson}.  Of particular note is the fact that this
formulation leads to Fock-like matrices which can be evaluated for
gradient corrected functionals without evaluating the Hessian of the
density.  For the high-spin open-shell states we use a spin-restricted
open-shell Kohn Sham (ROKS) procedure.  Some form of this approach has
apparently been used previously by Murray, Handy and
Amos\cite{murray-handy-amos} to study the ${}^3\mbox{B}_1$ state of
$\mbox{CH}_2$, but since they provide no details of the approach we
outline our implementation briefly below.  We adopt the notation of
Johnson et al.\cite{johnson}

\subsection{ROKS formalism}
One can view the Kohn-Sham equations as strictly analogous to the
Hartree-Fock equations except for the replacement of the
Hartree-Fock exchange operator with a local exchange-correlation
potential.  Given spin up and spin-down densities
$\rho_\alpha$ and $\rho_\beta$, evaluated on a
grid $\{ {\bf r}_j \}$ in space,
\begin{equation}
\rho_i({\bf r}_j)=\sum_{\mu\nu} P^i_{\mu\nu}\phi_\mu({\bf
r}_j)\phi_{\nu}({\bf r}_j),
\end{equation}
 where
$P^\alpha_{\mu\nu}$ and $P^{\beta}_{\mu\nu}$
are the familiar spin-up or spin-down density matrices and the $\phi$'s
are the elements of the basis set, one can express a general
functional of the density and its gradients as
\begin{equation}
f=f(\rho_\alpha,\rho_\beta,\gamma_{\alpha
\alpha},\gamma_{\alpha\beta},\gamma_{\beta\beta})
\end{equation}
where the gradient invariants
$\gamma$ are defined as
\begin{equation}
\gamma_{\alpha\alpha}=\left|\nabla\rho_\alpha\right|^2,
\gamma_{\alpha\beta}=\nabla\rho_\alpha \cdot \nabla\rho_\beta,
\gamma_{\beta\beta}=\left|\nabla\rho_\beta\right|^2 .
\end{equation}
The exchange-correlation energy can be written as
\begin{equation}
\label{eq:excnrg}
E_{XC}=\int f(\rho_\alpha,\rho_\beta,\gamma_{\alpha
\alpha},\gamma_{\alpha\beta},\gamma_{\beta\beta})  d\tau.
\end{equation}
In order to solve the Kohn-Sham equations, one forms a Fock-like matrix as
\begin{equation}
F^i_{\mu\nu}=h_{\mu\nu}+J_{\mu\nu}+F^{XCi}_{\mu\nu}
\end{equation}
where
\begin{equation}
\label{eq:fockel}
F^{XCi}_{\mu\nu}=\int \left\{
\frac{\partial f}{\partial\rho_i}\phi_\mu\phi_\nu + \left [
2\frac{\partial f}{\partial\gamma_{ii}}\nabla\rho_i + \frac{\partial
f}{\partial\gamma_{ij}}\nabla\rho_j\right]\cdot\nabla(\phi_\mu\phi_\nu)\right\}
d{\bf r},
\end{equation}
and $(i,j)\in\{(\alpha,\beta),(\beta,\alpha)\}.$ The matrices $h$ and
$J$ are the usual one-electron and Coulomb matrices, respectively.  In
spin-unrestricted Kohn-Sham (UKS) calculations, the $\alpha$ and
$\beta$ Fock matrices defined above are individually diagonalized and
the solutions iterated until self-consistency is achieved.  For the
spin-restricted high-spin open-shell calculations, we combine these
operators to obtain the analogue of a one-hamiltonian approach to
Hartree-Fock theory.  That is, if the orbitals are partitioned into
closed, open, and virtual blocks, the matrix
\begin{equation}\left(\begin{array}{ccc}
F_0&F_{CO}&F_{0}\\ &F_0 &F_{OV}\\ & &F_0
\end{array} \right)\end{equation}
is formed where
\begin{eqnarray}
F_0&=&\frac 12 (F^\alpha + F^\beta),\nonumber \\
F_{CO}&=&F^\beta,\\
F_{OV}&=&F^\alpha. \nonumber
\end{eqnarray}
This operator is identical to Hamilton and Pulay's one-hamiltonian
formulation of Hartree-Fock theory, except that the closed shell
exchange matrix ($K$) is replaced by $F^{XC\beta}$ and the open shell
exchange matrix with $F^{XC\alpha}-F^{XC\beta}$ (the unpaired
electrons are assigned alpha spin).  The Kohn-Sham matrix equations
are solved by the usual self-consistent techniques.  The open-shell
DIIS scheme of Hamilton and Pulay\cite{pulay1} was used to accelerate
convergence.  The Coulomb and one-electron terms are computed analytically.

\subsection{Functionals}
For convenience our program calculates exchange and correlation
functionals separately.  For most of the present work we have used the
gradient-corrected exchange functional of Becke\cite{becke-funct} and
the correlation functional of Lee, Yang and Parr \cite{LYP}.  We refer
to this combination as B-LYP.  For the copper atom we have also used
Becke's functional with no correlation (B-null).
\subsubsection{Becke Exchange}
Becke's exchange functional, which was designed to improve upon the
simple local density approximation (LDA) by yielding the correct
asymptotic behavior of the exchange energy, is given by
\begin{equation}
f_{B}(\rho_\alpha,\rho_\beta,\gamma_{\alpha\alpha},
\gamma_{\beta\beta})=\rho_\alpha^{4/3}g(x_\alpha)+\rho_\beta^{4/3}g(x_\beta)
\end{equation}
with
\begin{eqnarray}
g(x)&=&-\frac{3}{2}\left(\frac{3}{4\pi}\right)^{1/3}-
       \frac{bx^2}{1+6bx\sinh^{-1}x}\\
x_\alpha&=&\frac{\sqrt{\gamma_{\alpha\alpha}}}{\rho_\alpha^{4/3}}\\
x_\beta&=&\frac{\sqrt{\gamma_{\beta\beta}}}{\rho_\beta^{4/3}}.
\end{eqnarray}
Note that the expressions for $x_\alpha$ and $x_\beta$ given in
Reference~\citenum{johnson} contain a typographical error.  The parameter
$b$ is given by Becke\cite{becke-funct}, $b=0.0042$.

\subsubsection{Lee-Yang-Parr correlation}
The Lee-Yang-Parr (LYP) correlation functional is derived from the
correlation energy formula of Colle and Salvetti\cite{colle-salvetti},
derived from a consideration of short range effects in the
two-particle density matrix.  The functional itself, as transformed by
Miehlich et al.\cite{miehlich}, is given by
\begin{eqnarray}
\lefteqn{f_{LYP}(\rho_\alpha,\rho_\beta,\gamma_{\alpha\alpha},
         \gamma_{\alpha\beta},\gamma_{\beta\beta})=}\nonumber \\
& &-\frac{4a}{1+d\rho^{-1/3}}\frac{\rho_\alpha\rho_\beta}{\rho}\nonumber\\
& &-2^{11/3}\frac{3}{10}(3\pi^2)^{2/3}ab\omega(\rho)\rho_\alpha\rho_\beta
         (\rho_\alpha^{8/3}+\rho_\beta^{8/3})\\
& &+\frac{\partial f_{LYP}}{\partial
\gamma_{\alpha\alpha}}\gamma_{\alpha\alpha} +\frac{\partial f_{LYP}}{\partial
\gamma_{\alpha\beta}}\gamma_{\alpha\beta} +\frac{\partial f_{LYP}}{\partial
\gamma_{\beta\beta}}\gamma_{\beta\beta} \nonumber
\end{eqnarray}
with
\begin{eqnarray}
\frac{\partial f_{LYP}}{\partial\gamma_{\alpha\alpha}}&=&
            -ab\omega(\rho)\left[\frac
19 \rho_\alpha\rho_\beta\left\{1-3\delta(\rho) -[\delta(\rho)-11]\frac
{\rho_\alpha}{\rho}\right\}-\rho_\beta^2\right],\\
\frac{\partial
f_{LYP}}{\partial\gamma_{\alpha\beta}}&=&-ab\omega(\rho)\left\{\frac
19 \rho_\alpha\rho_\beta[47-7\delta(\rho)]-\frac 43\rho^2\right\},\\
\frac{\partial f_{LYP}}{\partial\gamma_{\beta\beta}}&=&-ab\omega(\rho)
       \left[\frac
19 \rho_\alpha\rho_\beta\left\{1-3\delta(\rho)-[\delta(\rho)-11]\frac
{\rho_\beta}{\rho}\right\}-\rho_\alpha^2\right],\\
\omega(\rho)&=&\frac{e^{-c\rho^{-1/3}}}{1+d\rho^{-1/3}} \rho^{-11/3},\\
\delta(\rho)&=&c\rho^{-1/3}+\frac{d\rho^{-1/3}}{1+d\rho^{-1/3}}.
\end{eqnarray}
The constants $a$,$b$, $c$, and $d$ used were those used by Miehlich
et al.\cite{miehlich}, $a=0.04918$, $b=0.132$, $c=0.2533$, and
$d=0.349$.

\subsection{Grids and Integration Scheme}
The evaluation of the exchange-correlation contribution to the Fock
matrix (Eq.~\ref{eq:fockel}) and the total energy
(Eq.~\ref{eq:excnrg}) is done by a numerical quadrature.  In this
paper we report atomic results but we have implemented the more
general partitioning scheme of Becke\cite{becke-grid}.  For each atom
in the molecule an atomic grid is generated consisting of concentric
spherical shells centered on the atom.  The shells are given radii
according to the Euler-MacLaurin scheme described by Murray, et
al.\cite{murray}; each spherical shell has points and weights
distributed on it according to the formul\ae\ of Lebedev
\cite{lebedev}.  When the atomic grids are assembled into a
molecular grid the weights are adjusted to partition the space into
``fuzzy Voronoi polyhedra''.  For these atomic calculations the
Lebedev angular grid of order 9 was used and is more than sufficient
to integrate exactly all spherical harmonics which may appear in the
solutions.  The radial grid consisted of 100 points, and the ``radial
maximum'' parameter chosen according to Slater's rules.  In
particular, we used values of 4.59, 4.38, 4.20, 4.01, 3.82, 3.69,
3.53, 3.40, 3.27 and 3.16$a_0$ for the atoms Sc--Zn, respectively.
The accuracy of the radial grid was tested on copper by comparing the
results using both larger (150 and 200 radial points) and smaller (50
and 75 points) grids.  The use of 100 radial points yields stability
in the total energy (in a.u.) to four digits past the decimal point;
increasing the grid to 150 points changes only the fifth digit, and
using only 50 points changes the second digit.

\subsection{Basis Sets}

The Kohn-Sham orbitals were expanded in a primitive Cartesian Gaussian
basis set of dimension (14s9p6d).  This consists of Wachters'
primitive set\cite{wachters} augmented with the diffuse $d$ function
recommended by Hay\cite{hay-dif-d}.  All six Cartesian components of
the $d$ functions were retained.  Since Wachters' primitive basis was
optimized for Hartree-Fock calculations, and the detailed shape of the
Kohn-Sham orbitals can, in principle, be different, some attention was
paid to the nature of the Kohn-Sham coefficients as an indication that
the primitive basis should be modified.  In general it appears that
the Wachters/Hay set is adequate to expand the Kohn-Sham orbitals as
well.  In particular, most orbitals were described by two or more
primitives with coefficients of $\sim 0.5$.  There were a few
exceptions which might warrant further study.  For example, whereas
the Ti $2s$ Hartree-Fock orbital involves primitives 9 and 10 with
coefficients of 0.5 and 0.6, the Ti $2s$ B-LYP orbital wants to be
somewhat more diffuse as reflected by coefficients for primitives 9
and 10 of 0.2 and 1.0.  If one makes an even-tempered plot of the Ti
$s$-space, it is apparent that there is a ``gap'', or missing
$s$-function in the sequence which reflects the 2$s$/3$s$ shell
structure.  An additional function would presumably improve the
description of this core orbital and reduce the total energy somewhat,
but should have little effect on the valence electron distribution or
relative energy differences.  Additional examples of this sort occur
for atoms on the right side of the series.

We have also performed some initial investigations on contracted basis
sets appropriate for use in the first transition series.  These
results are discussed in Section 3.3.

\subsection{Convergence issues}
For many of the cases described below we found that the simple ROKS
procedure failed to converge.  In most of these cases the cause of the
failure was a tendency of the unpaired electrons to occupy different
components of the $3d$ orbital from one iteration to the next.  In
most cases all that was required to avoid this was to employ symmetry
constraints; for example, for the $3d^24s$ state of scandium we used
the octahedral point group with inversion, and required that the two
singly occupied $d$ orbitals be those of the $E_g$ irreducible
representation, namely $d_{2z^2-x^2-y^2}$ and $d_{x^2-y^2}$.  An
initial guess with the proper symmetry was formed by permuting the
default initial guess vectors.  At each step the Fock matrix was
diagonalized in the symmetrized orbital basis, and the occupied
orbitals chosen so that the number of occupied orbitals in each
irreducible representation remained the same as for the initial guess.
Enforcing $O_h$ symmetry also prevents the $s$ orbitals from mixing
with the $d_{z^2}$ orbital.

While using symmetry constraints was sufficient to ensure convergence
for most states, there were a few stubborn cases, such as
$\mbox{Sc}^+(3d4s)$.  In these cases there were no symmetry groups
which could both prevent mixing of the $s$ and $d$ orbitals and still
prevent unpaired $d$ electrons from moving between degenerate
symmetry-equivalent $d$ orbitals.  We resolved this problem by
employing, for those cases where it was necessary, a maximum overlap
condition in addition to the symmetry constraints: orbitals were
filled according to the maximum overlap with the initial guess.  The
cases for which this was used were: $\mbox{Sc}^+(3d4s)$,
$\mbox{V}(3d^44s)$, $\mbox{Cr}(3d^6)$, $\mbox{Cr}(3d^44s^2)$,
$\mbox{Cr}^+(3d^44s)$, $\mbox{Fe}(3d^64s^2)$, $\mbox{Fe}^+(3d^64s)$,
and $\mbox{Co}(3d^9)$.

\section{Results}
\subsection{Ionization and excitation energies}

\begin{figure}[h]
\leavevmode
\epsffile{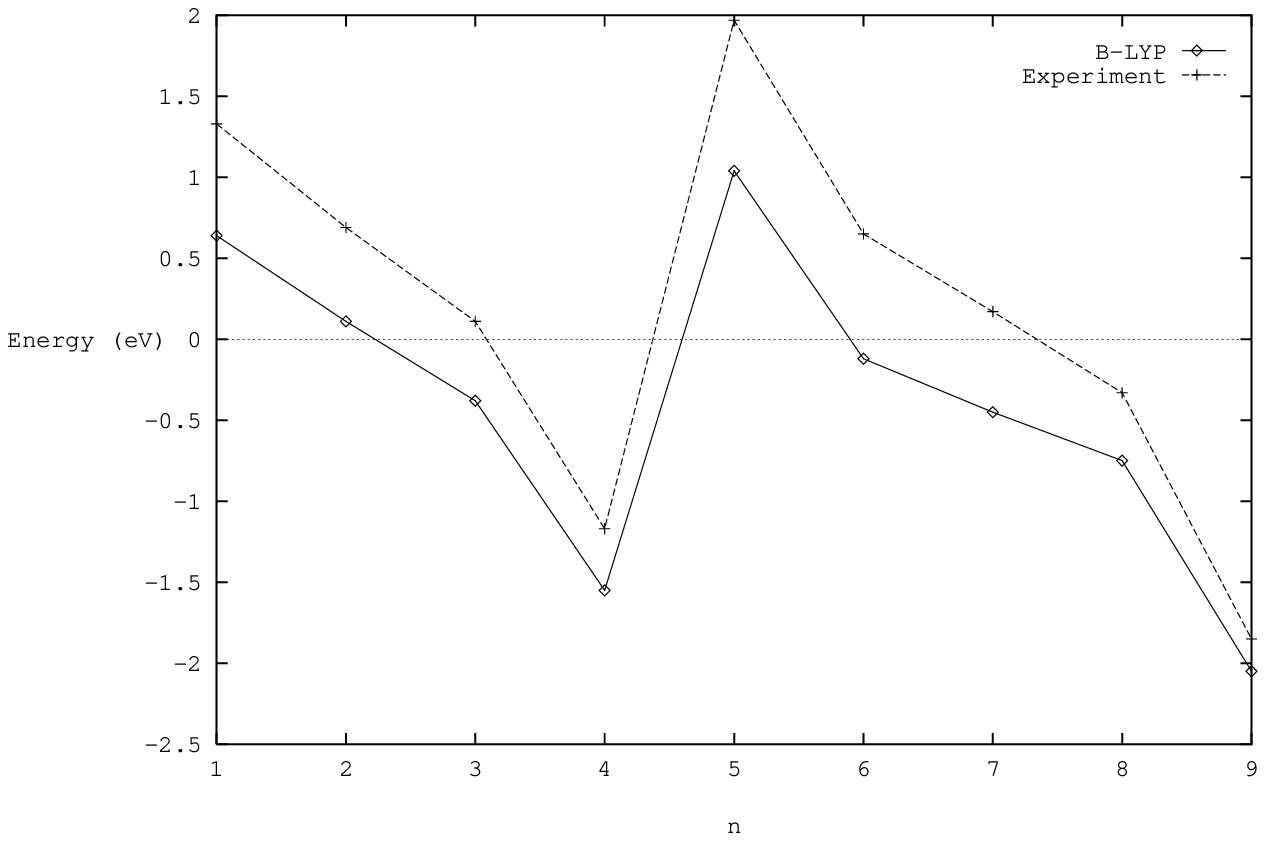}
\caption{
Plot of experimental and B-LYP values for the interconfigurational
energy between the $d^ns^2$ and $d^{n+1}s$ states for the first
transition series.  Diamonds are B-LYP values, plusses are
experimental values with relativistic corrections.}
\label{gf:eic}
\end{figure}

\begin{figure}[h]
\leavevmode
\epsffile{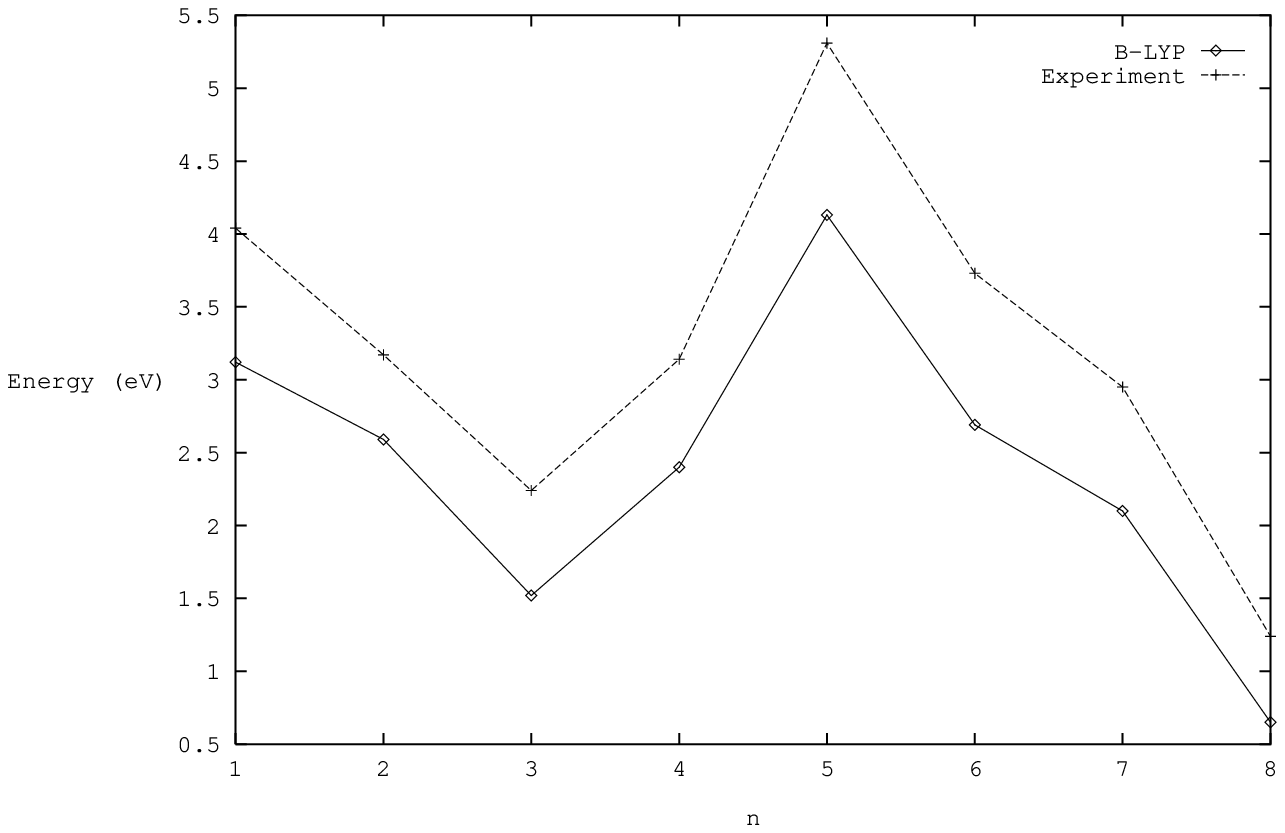}
\caption{
Plot of experimental and B-LYP values for the interconfigurational
energy between the $d^ns^2$ and $d^{n+2}$ states for the first
transition series.  Diamonds are B-LYP values, plusses are
experimental values with relativistic corrections.}
\label{gf:eic4}
\end{figure}

\begin{figure}[h]
\leavevmode
\epsffile{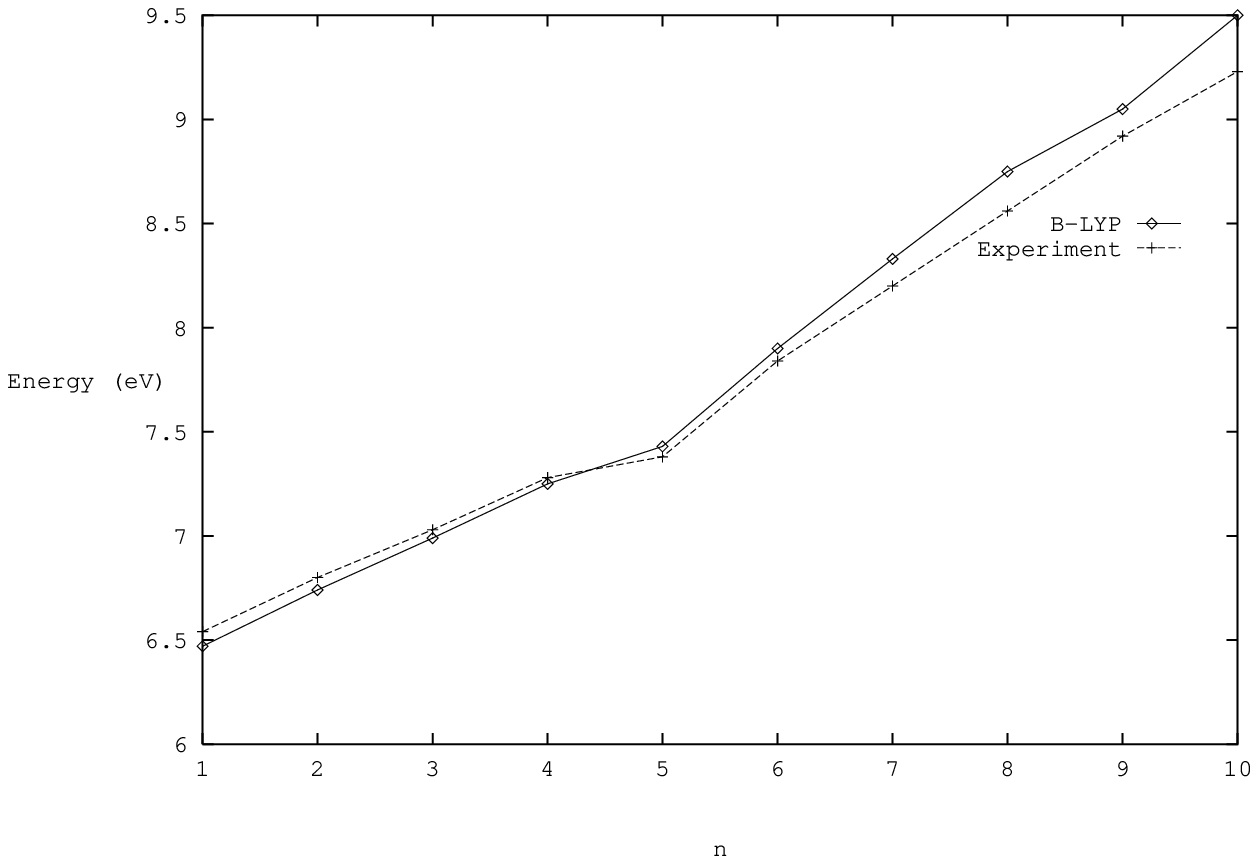}
\caption{
Plot of experimental and B-LYP values for the interconfigurational
energy between the $d^ns^2$ and $d^ns$ states for the first
transition series.  Diamonds are B-LYP values, plusses are
experimental values with relativistic corrections.}
\label{gf:eic2}
\end{figure}

\begin{figure}[h]
\leavevmode
\epsffile{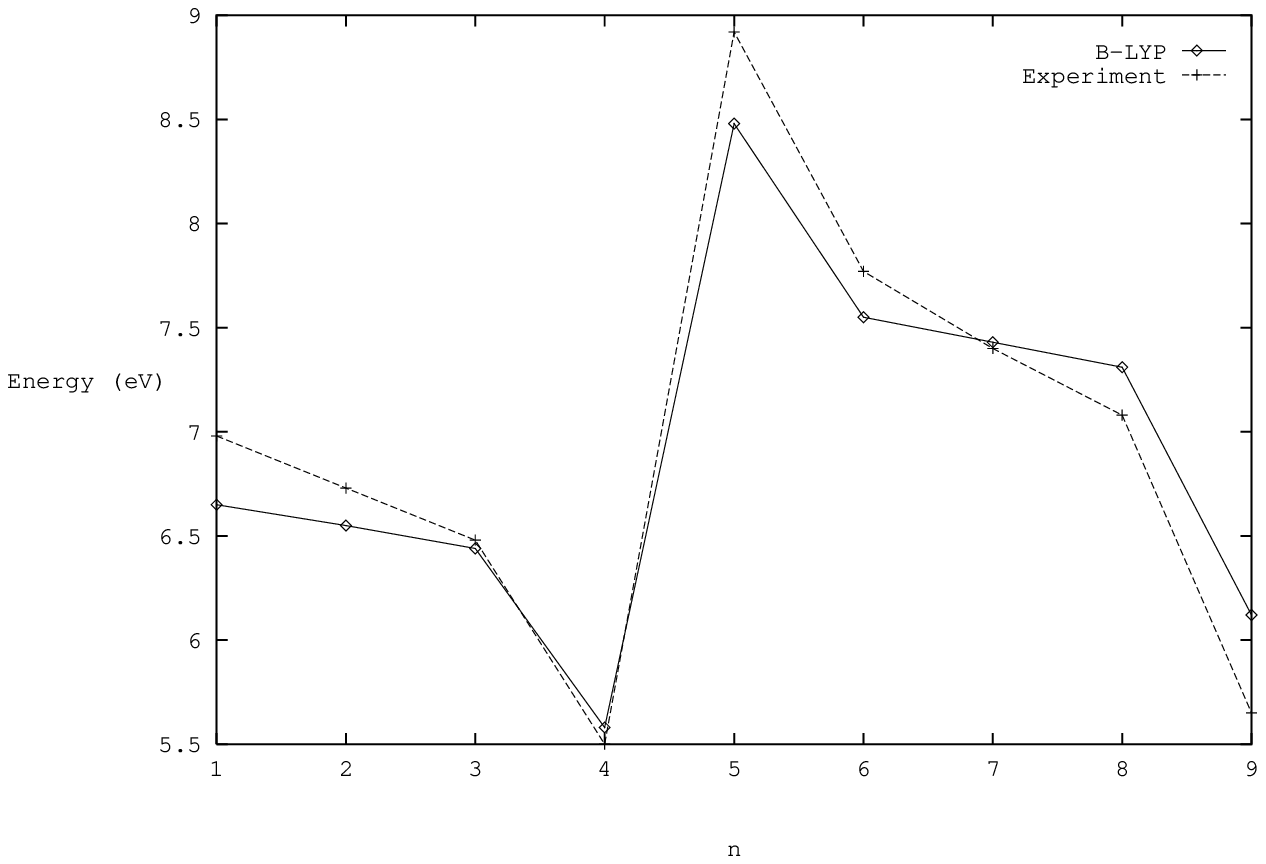}
\caption{
Plot of experimental and B-LYP values for the interconfigurational
energy between the $d^ns^2$ and $d^{n+1}$ states for the first
transition series.  Diamonds are B-LYP values, plusses are
experimental values with relativistic corrections.}
\label{gf:eic3}
\end{figure}

The results of the B-LYP calculations of the ionization and excitation
energies of the first transition series are given in
Tables~\ref{tb:metals} and~\ref{tb:metals2}.
The experimental values corrected for relativistic contributions are
given in the final column of Tables~\ref{tb:metals}
and~\ref{tb:metals2}.  These values are simply the experimental
results with an estimate of the differential relativistic energy
subtracted out, as was done by Raghavachari and
Trucks\cite{raga-exciting,raga-ions}.  The estimate is based on
directly computed relativistic contributions to the Hartree-Fock
energies\cite{rlm-hay-xnmet}, scaled by an estimate for the effects of
electron correlation.  Figures~\ref{gf:eic}--\ref{gf:eic3} compare the
B-LYP results with these corrected experimental values.

Perusal of Tables~\ref{tb:metals} and~\ref{tb:metals2} shows that the
B-LYP approximation gives a generally reliable picture of the relative
stability of the states of the first transition series.  Consider
first the $d^ns^2\rightarrow d^{n+1}s$ excitation energy shown in
Figure~\ref{gf:eic}.  Note that the general ``sawtooth'' behavior in
the excitation energy is faithfully reproduced by the restricted
open-shell B-LYP method.  Nevertheless, the ground state of the atom
is predicted incorrectly in a few cases: vanadium, where B-LYP
predicts the $d^4s$ state to be more stable than $d^3s^2$; iron, where
B-LYP yields $d^7s$ lower than $d^6s^2$; and cobalt, where B-LYP
places $d^8s$ lower than $d^7s^2$.  A general tendency of B-LYP to
favor $d^{n+1}s$ over $d^ns$ configurations is apparent in
Figure~\ref{gf:eic}.  The $d^{n+1}s$ states are generally predicted to
be too stable by $\sim 0.5\mbox{eV}$.  In V, Co and Fe the
experimental splitting between $d^{n+1}s$ and $d^ns^2$ states is of
this order or smaller, and so the bias leads to an incorrect ordering.
This tendency to favor $d^{n+1}s$ states is common to the LDA as well.
Harris and Jones\cite{harris-jones} found a bias towards $d^{n+1}s$ of
$\sim 1\mbox{eV}$.
Theory and experiment are compared for the $d^ns^2\rightarrow d^{n+2}$
excitation in Figure~\ref{gf:eic4}.  Again, the qualitative features
of the trends are reproduced well, but the configurations rich in
$d$-electrons (or poor in $s$-electrons) are consistently predicted to
be too stable, in this case by $\sim 0.8\mbox{eV}$.

Figure~\ref{gf:eic2} shows the results for the ionization potential
$d^ns^2\rightarrow d^ns$, a transition in which the number of
$d$-electrons remains constant.  Here, the B-LYP approximation is in
much better agreement with experiment.  This would suggest that the
errors in the excitation energies of the neutral arise primarily from
a tendency to overbind the $d$-electrons, and that the ionization
potential of the $s$-electron is roughly correct.  This conclusion is
generally consistent with the results for the $d^ns^2\rightarrow
d^{n+1}$ ionization potential plotted in Figure~\ref{gf:eic3}.  Most
are underestimated consistent with a bias toward configurations rich
in $d$-electrons, although this rule of thumb would not predict the
overestimates which occur at the far right of the series.

\begin{figure}[h]
\leavevmode
\epsffile{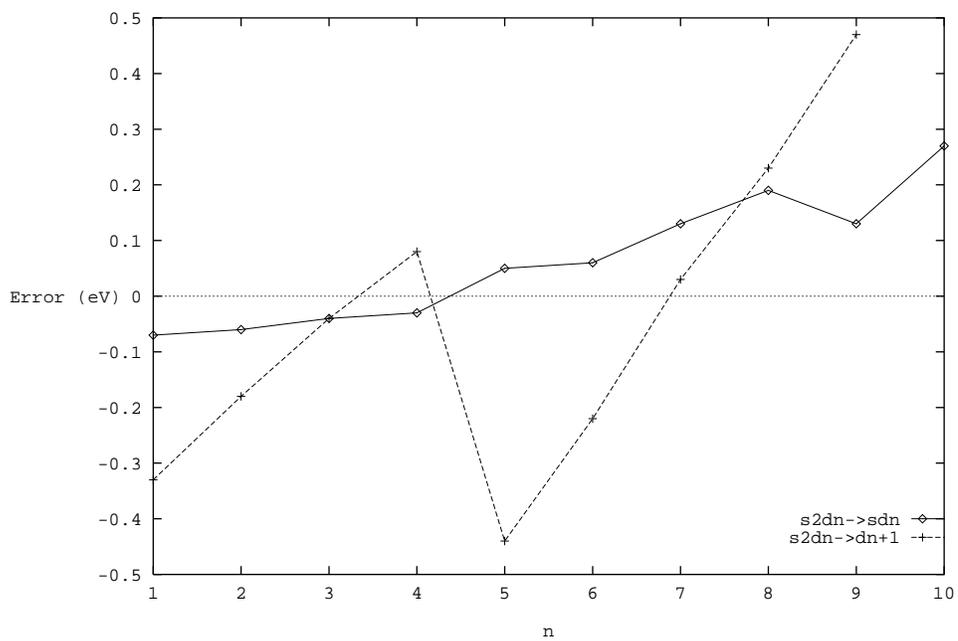}
\caption{
Plot of difference between experimental values and B-LYP values of
ionization potentials for the first row transition metals.
Diamonds are the $s^2d^n\rightarrow sd^n$ ionization potential,
plusses are $s^2d^n\rightarrow d^{n+1}$.
}
\label{gf:ions}
\end{figure}

The mean absolute errors in the B-LYP approximation for the various
excitation energies are given in the final column of
Table~\ref{tb:compare-hfetc} , and the individual errors plotted in
Figures~\ref{gf:excite} and \ref{gf:ions}.  For purposes of
comparison, Table~\ref{tb:compare-hfetc} also contains the mean
absolute errors as computed by Raghavachari and Trucks at the
Hartree-Fock, M\o ller-Plesset, and QCISD(T) levels of approximation.
For the $d^ns^2\rightarrow d^ns$ ionization potential, the mean
absolute B-LYP error is only 0.10 eV, and compares favorably with both
the MP4(0.10eV) and QCISD(T)(0.09eV) results.  The B-LYP error is
somewhat larger (0.21eV) for the $d^ns^2\rightarrow d^{n+1}$
ionization, but is again competitive with the QCISD(T) approximation
(0.14eV), and clearly superior to MP4.  B-LYP does not perform as well
for the excitation energies of the neutral, with mean absolute errors
of 0.51eV and 0.76 for the $d^ns^2\rightarrow d^{n+1}s$ and
$d^ns^2\rightarrow d^{n+2}$ transitions, respectively.  The former can
be compared with the QCISD(T) mean absolute error of 0.12eV.  The
overall mean B-LYP errors, it should be emphasized, are significantly
smaller than the error from Hartree-Fock calculations, which can be as
large as 4eV for the $d^ns^2\rightarrow d^{n+2}$ excitation.  QCISD(T)
results were not reported for this excitation.

Thus far we have compared the gradient-corrected B-LYP DFT results
with Hartree-Fock based approximations.  It is also instructive to
compare the B-LYP results with other gradient-corrected DFT
functionals, such as the Becke-Perdew (BP) variant.  Table~\ref{tb:bp}
compares the mean absolute errors in B-LYP ionization potentials with
the recent results of Ziegler and Li\cite{ZiegLi}, who examined both
the Becke-Perdew approximation and the LDA (Slater exchange and
Vosko-Wilk-Nusair correlation functionals).  While it should be kept
in mind that this is not a completely fair comparison of the
functionals since Ziegler and Li used the spin-unrestricted version
of Kohn-Sham theory and basis sets different from ours, it is
interesting to note that both the B-LYP and BP functionals give quite
similar results.  It is also interesting that the LDA, while in worse
agreement with experiment than either of the gradient-corrected
approaches, performs rather well.  In this series of ionization
potentials, the gradient corrections do not appear to give
dramatically improved results.

This observation is also in accord with the recent work of Kutzler and
Painter\cite{k-p}.  For the $s$-ionization potentials, e.g., they find
mean absolute errors of 0.22eV and 0.39eV for the gradient-corrected
functionals of Langreth, Mehl and Hu (LMH)\cite{LMH} and the
generalized-gradient-approximation (GGA) of Perdew and Yue\cite{p-u},
respectively.  These can be compared with the B-LYP error of 0.10eV,
the Becke-Perdew result of 0.16eV, and the LDA error of 0.28eV.

For the $s\rightarrow d$ excitation energies in the neutral species,
the LDA meann absolute error is $\sim 0.85\mbox{eV}$\cite{k-p}.  The
B-LYP error found in the present work (0.51eV) is comparable to the
GGA result ($\sim 0.6\mbox{eV}$) of Kutzler and Painter, and
significantly better than the LMH error ($\sim 1.2\mbox{eV}$).

\subsection{Analysis of B-LYP functional}
Although
the B-LYP error is large for the $d^ns^2\rightarrow d^{n+1}s$
excitation energy, the fact that the qualitative features of the
experimental trends are well reproduced (cf. Figure~\ref{gf:eic})
suggests that this might be traced to a systematic error which could
be improved with future functionals.  An immediate question, then, is
where does the error reside, in the exchange or correlation functional?
The origin of the error is not as easy to pin down as one might think
at first.  Consider the $d^9s^2\rightarrow d^{10}s$ transition in the
copper atom.  The difference in the exact exchange contribution
between these two states can be extracted from Hartree-Fock
calculations on the two.  It turns out to be -5.62eV (see
Table~\ref{tb:pickapart}).  The differential exchange energy from
either the B-null or B-LYP calculation is very similar (-5.69 and
-5.74eV, respectively).  The difference in the correlation energy
between the two states can be inferred from the Hartree-Fock
calculations and the experimental results; it is -1.54eV.  The
correlation energy from the B-LYP calculations is only -0.16eV.  One
thus might expect a rather large total B-LYP error of +1.26eV (-0.12eV
from the exchange and +1.38eV from the correlation energy error).  The
actual error in the B-LYP calculation for this case is -0.2eV.  The
reason for this is that the B-LYP one-electron and Coulomb
contributions are different from the Hartree-Fock values.  The
self-consistency aspect of the calculation makes a direct examination
of the error difficult.

In order to make a more direct comparison, we also examined the
results of the B-LYP functional being applied to the Hartree-Fock
density.  These numbers are also displayed in
Table~\ref{tb:pickapart}.  The differential Becke exchange energy
(-6.99eV) is now quite different from the B-LYP value (-5.74eV) and in
significant disagreement with the Hartree-Fock value (-5.62eV).  The
differential LYP correlation energy is hardly changed (-0.16 and
-0.18eV) and still significantly underestimated.  Thus it appears that
there are rather large errors in both functionals.  The total
B-LYP(HF) prediction is in rather good agreement with experiment, but
this comes about because the overestimate of the exchange energy tends
to cancel the underestimate of the correlation energy.
Table~\ref{tb:pickapart} provides data for a similar analysis of the
other states of interest in the copper atom.  They are consistent with
the conclusion reached above.  Thus it appears that for Cu, the LYP
functional underestimates the correlation energy, and that the Becke
functional overestimates the exchange energies.  The total
exchange-correlation energy is reproduced rather accurately, and both
the B-LYP and B-LYP(HF) approaches are in good agreement with
experiment.

\begin{figure}[h]
\leavevmode
\epsffile{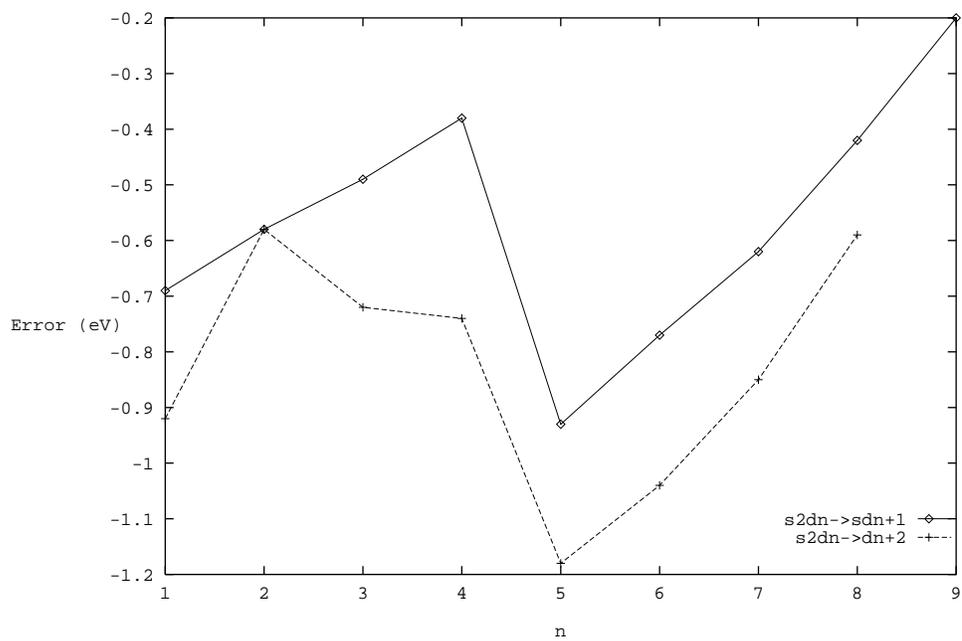}
\caption{
Plot of difference between experimental values and B-LYP values of
excitation energies for the first row transition metals.
Diamonds are the $s^2d^n\rightarrow sd^{n+1}$ excitation energy,
plusses are $s^2d^n\rightarrow d^{n+2}$.
}
\label{gf:excite}
\end{figure}

The analysis and the discussion above leads one to conclude that the
accuracy in the final B-LYP result arises from a rather fortuitous
cancellation of errors in the exchange and correlation functionals.
However, there appear to be some systematics to the error, as a study
of Figures~\ref{gf:excite} and \ref{gf:ions} demonstrate.  With the
exception of the $d^ns^2\rightarrow d^ns$ ionization potential, in
which the error is roughly linear in $n$, the curves typically exhibit
a sawtooth behavior, in which the error at $n=5$(Mn) suddenly becomes
more severe.  At first glance, one might attempt to associate the
sudden change in the error for $d^ns^2\rightarrow d^{n+1}s$ which
occurs between Cr and Mn ($n=4$ and $n=5$) to the sudden appearance of
doubly occupied $d$-subshells in the $d^6s$ state of Mn.  For example,
the error in the $d^4s^2\rightarrow d^5s$ excitation energy is $\sim
-0.4\mbox{eV}$ in Cr, abruptly increasing to $\sim -0.9\mbox{eV}$ for
the $d^5s^2 \rightarrow d^6s$ excitation in Mn.  Since the $d^6s$
state is predicted to be too stable relative to $d^5s^2$, one might
conclude that the correlation functional overestimates the intra-pair
$d$-$d$ correlation energy.  However, this argument would predict a
similar increase in the error for the $d^4s^2\rightarrow d^6$
excitation in Cr.  Figure~\ref{gf:excite} shows that this is not the
case, and that the increase again occurs at Mn, this time in the
$d^5s^2\rightarrow d^7$ excitation energy.  Similarly, if one argues
that the error is associated with the exchange functional and the
``special'' nature of the half-filled $d^5$ shell, then one is
hard-pressed to explain why a similar jump in the error is not
apparent in the $d^3s^2\rightarrow d^5$ excitation energy of Ti.  At
present we do not understand this behavior and further work is clearly
warranted for this problem.

\subsection{Basis Set Contractions}

In order to test the functionals, we have attempted to eliminate many
of the uncertainties associated with basis set incompleteness by using
the fully uncontracted $(14s9p6d)$ basis of Wachters/Hay.  For
applications to molecules this basis is too large, and should be
appropriately contracted.  An obvious approach would be to use a
general contraction scheme based on the Kohn-Sham orbitals obtained in
this work.  One might expect, however, that different contractions
would be necessary for different variants of DFT; e.g., a set of
contractions appropriate for LDA calculations, a different set for
B-LYP, etc.  For our initial investigation, we decided to test a
contraction scheme based on the Hartree-Fock orbitals.  We therefore
contracted the $(14s9p6d)$ primitive basis into a [6s5p3d] basis using
the general contraction scheme of Raffenetti\cite{gen-contract} and
the Hartree-Fock coefficients of Wachters\cite{wachters}.
Specifically, the inner parts of the $1s$, $2s$, $2p$, $3s$, $3p$ and
$3d$ orbitals were contracted via the HF coefficients, and the
remaining, more diffuse primitives in each space were left free.
These results are also presented in Tables~\ref{tb:metals} and
\ref{tb:metals2}.
While the absolute energies were affected as expected, the excitation
and ionization energies were generally changed by at most 0.02eV.  The
exceptions to this behavior occur for the Ni and Cu atoms, where some
of the relative energies changed by as much as 0.12 eV.

\section{Conclusions}

We have examined the predictions of spin-restricted Kohn-Sham theory
as regards the excitation and ionization energies of the members of
the first transition series using the gradient-corrected density
functionals of Becke and Lee, Yang and Parr.

First of all, it is important to note that the qualitative features of
the trends in excitation and ionization energies
(Figs.~\ref{gf:eic}--\ref{gf:eic3}) are faithfully reproduced with a
spin-restricted formulation of Kohn-Sham theory.  It has sometimes
been suggested in the literature that a spin-unrestricted formulation
is necessary to reproduce, e.g., the break in the $d^ns^2\rightarrow
d^{n+1}s$ excitation energy which occurs at Cr (Fig.~\ref{gf:eic}).
This is clearly not the case.  We believe this to be important, for a
spin-restricted approach  has the advantage that the solutions are
eigenfunctions of spin and the uncontrolled spin contamination which
freqently occurs in unrestricted Hartree-Fock (UHF) or Kohn-Sham (UKS)
calculations on transition metal complexes is thereby avoided.  We
suspect that ROKS shares the disadvantages of ROHF:  simple bonds will
not always dissociate properly, molecules in which the qualitative
electronic structure is best viewed as biradical in character may not
be treated well, etc. In the present context, one might expect a
significant spin polarization in the unrestricted formalism for those
atomic states with a large number of unpaired $d$-electrons.  While
this can certainly lead to total energies which are significantly
lower than the spin-restricted energies, it is interesting to note
that the relative energies calculated by the ROKS procedure in this
work are in close agreement with the UKS results of Ziegler and Li.  A
direct comparison of ROKS {\em vs.} UKS is made difficult by the fact
that Ziegler and Li used the Becke-Perdew functional whereas we
examined Becke-LYP.  The close agreement suggests, however, that
neither the spin-polarization effects nor the different functionals
used cause significant differences for the predictions of the
ionization potentials of the first transition series.

The quantitative agreement with experiment is also encouraging.
Ionization energies are observed to be in good agreement with
experiment, with mean absolute errors of 0.10eV for the
$d^ns^2\rightarrow d^ns$ ionization, and a mean absolute error of
0.21eV for $d^ns^2\rightarrow d^{n+1}$.  These errors are much smaller
than those obtained by Hartree-Fock, and compare favorably with the
QCISD(T) results of Raghavachari and Trucks, which have mean absolute
errors of 0.09 and 0.14eV, respectively.  The ROKS B-LYP errors in the
excitation energies are larger, 0.51eV for the $d^ns^2\rightarrow
d^{n+1}s$ excitation, as compared with an average absolute error of
0.12eV from the QCISD(T) calculations.  The B-LYP approximation, like
other DFT variants, consistently places the $d^{n+1}s$ states too low
compared with $d^ns^2$.

A comparison of the results using the primitive (14s9p6d) basis of
Wachters versus a (6s5p3d) general contraction based on Hartree-Fock
coefficients shows that the contracted basis is generally in excellent
agreement with the fully uncontracted basis.  This demonstrates that
the description of the potential in the valence region of the atom due
to the core electrons is essentially the same in B-LYP and
Hartree-Fock calculations.  This suggests that the relativistic
effective core potentials developed for the Hartree-Fock problem may
be applicable to the DFT methods as well.

In summary, the spin-restricted Kohn-Sham calculations using the B-LYP
functional look promising for calculations on transition metal
complexes.  The atomic errors are significantly smaller than the HF
and MP approximations, particularly for those elements to the right of
the row where the MP series is slow to converge.  While not as
accurate as the coupled-cluster techniques for this row, e.g.
QCISD(T), B-LYP achieves a reasonable level of accuracy at a much
reduced level of effort.  Although it is clear that improved
functionals are needed for quantitative accuracy, we expect the B-LYP
functional, its variants or descendants to be increasingly used for
electronic structure calculations on transition metal complexes.

\section{Acknowledgments}
This work was sponsored by the U.S. Department of Energy.
\begin{table}[ht]
\begin{center}
\begin{tabular}{lrrrrrrr}\hline
Atom&State&Energy (au)&$\Delta E$ (eV)&Energy (au)&$\Delta E$ (eV)&Rel. Corr.\\
    &     &$(14s9p6d)$&               &$[6s5p3d]$  & &Exp.
\cite{raga-exciting,raga-ions}\\\hline\hline
Sc           &$3d4s^2({}^2D)$  & -760.6369 & 0.00 & -760.6310 & 0.00 &      \\
             &$3d^24s({}^4F)$  & -760.6132 & 0.64 & -760.6071 & 0.64 & 1.33 \\
             &$3d^3({}^4F)$    & -760.5224 & 3.12 & -760.5166 & 3.11 & 4.04 \\
      \hline
$\mbox{Sc}^+$&$3d4s({}^3D)$    & -760.3992 & 6.47 & -760.3932 & 6.47 & 6.54 \\
             &$3d^2({}^3F)$    & -760.3925 & 6.65 & -760.3866 & 6.65 & 6.98 \\
\hline\hline
Ti           &$3d^24s^2({}^3F)$& -849.3803 & 0.00 & -849.3729 & 0.00 &      \\
             &$3d^34s({}^5F)$  & -849.3764 & 0.11 & -849.3691 & 0.10 & 0.69 \\
             &$3d^4({}^5D)$    & -849.2851 & 2.59 & -849.2781 & 2.58 & 3.17 \\
\hline
$\mbox{Ti}^+$&$3d^24s({}^4F)$  & -849.1326 & 6.74 & -849.1252 & 6.74 & 6.80 \\
	     &$3d^3({}^4F)$    & -849.1397 & 6.55 & -849.1326 & 6.54 & 6.73 \\
\hline\hline
V	     &$3d^34s^2({}^4F)$& -943.9282 & 0.00 & -943.9200 & 0.00 &      \\
	     &$3d^44s({}^6D)$  & -943.9422 &-0.38 & -943.9341 &-0.38 & 0.11 \\
	     &$3d^5({}^6S)$    & -943.8724 & 1.52 & -943.8646 & 1.50 & 2.24 \\
 \hline
$\mbox{V}^+$ &$3d^34s({}^5F)$  & -943.6711 & 6.99 & -943.6628 & 7.00 & 7.03 \\
	     &$3d^4({}^5D)$    & -943.6913 & 6.45 & -943.6833 & 6.44 & 6.48 \\
\hline\hline
Cr           &$3d^44s^2({}^5D)$&-1044.4196 & 0.00 &-1044.4100 & 0.00 &      \\
	     &$3d^54s({}^7S)$  &-1044.4767 &-1.55 &-1044.4673 &-1.56 &-1.17 \\
	     &$3d^6({}^5D)$    &-1044.3314 & 2.40 &-1044.3213 & 2.41 & 3.14 \\
 \hline
$\mbox{Cr}^+$&$3d^44s({}^6D)$  &-1044.1533 & 7.25 &-1044.1442 & 7.23 & 7.24 \\
	     &$3d^5({}^6S)$    &-1044.2144 & 5.58 &-1044.2052 & 5.57 & 5.46 \\
\hline\hline
Mn  	     &$3d^54s^2({}^6S)$&-1151.0244 & 0.00 &-1151.0132 & 0.00 &      \\
	     &$3d^64s({}^6D)$  &-1150.9863 & 1.04 &-1150.9747 & 1.05 & 1.97 \\
	     &$3d^7({}^4F)$    &-1150.8728 & 4.13 &-1150.8614 & 4.13 & 5.31 \\
 \hline
$\mbox{Mn}^+$&$3d^54s({}^7S)$  &-1150.7514 & 7.43 &-1150.7401 & 7.43 & 7.38 \\
	     &$3d^6({}^5D)$    &-1150.7126 & 8.48 &-1150.7007 & 8.50 & 8.92 \\
\hline\hline
Fe	     &$3d^64s^2({}^5D)$&-1263.7208 & 0.00 &-1263.7145 & 0.00 &      \\
	     &$3d^74s({}^5F)$  &-1263.7253 &-0.12 &-1263.7189 &-0.12 & 0.65 \\
	     &$3d^8({}^3F)$    &-1263.6219 & 2.69 &-1263.6154 & 2.70 & 3.73 \\
 \hline
$\mbox{Fe}^+$&$3d^64s({}^6D)$  &-1263.4303 & 7.90 &-1263.4240 & 7.90 & 7.84 \\
	     &$3d^7({}^4F)$    &-1263.4434 & 7.55 &-1263.4369 & 7.55 & 7.77 \\
\hline\hline
\end{tabular}
\end{center}
\caption{ Results for B-LYP runs on the transition metals.}
\label{tb:metals}
\end{table}
\begin{table}[ht]
\begin{center}
\begin{tabular}{lrrrrrrr}\hline
Atom&State&Energy (au)&$\Delta E$ (eV)&Energy (au)&$\Delta E$ (eV)&Rel. Corr.\\
    &     &$(14s9p6d)$&               &$[6s5p3d]$ &
&Exp.\cite{raga-exciting,raga-ions}\\\hline\hline
Co     &$3d^74s^2({}^4F)$   &-1382.7912 & 0.00 &-1382.7831 & 0.00 &      \\
       &$3d^84s({}^4F)$     &-1382.8076 &-0.45 &-1382.7996 &-0.45 & 0.17 \\
   &$3d^9({}^2D)$       &-1382.7139 & 2.10 &-1382.7057 & 2.11 & 2.95 \\\hline
$\mbox{Co}^+$&$3d^74s({}^5F)$   &-1382.4850 & 8.33 &-1382.4769 & 8.33 & 8.20 \\
 &$3d^8({}^3F)$  &-1382.5179 & 7.43 &-1382.5098 & 7.43 & 7.40 \\\hline \hline
Ni         &$3d^84s^2({}^3F)$   &-1508.3439 & 0.00 &-1508.3289 & 0.00 &      \\
	   &$3d^94s({}^3D)$     &-1508.3716 &-0.75 &-1508.3592 &-0.82 &-0.33 \\
   &$3d^{10}({}^1S)$    &-1508.3199 & 0.65 &-1508.3060 & 0.62 & 1.24 \\\hline
$\mbox{Ni}^+$&$3d^84s({}^4F)$   &-1508.0225 & 8.75 &-1508.0079 & 8.73 & 8.56 \\
 &$3d^9({}^2D)$  &-1508.0751 & 7.31 &-1508.0612 & 7.28 & 7.08 \\\hline\hline
Cu     &$3d^94s^2({}^2D)$   &-1640.5003 & 0.00 &-1640.4797 & 0.00 &      \\
&$3d^{10}4s({}^2S)$  &-1640.5758 &-2.05 &-1640.5586 &-2.15 &-1.85 \\\hline
$\mbox{Cu}^+$&$3d^94s({}^3D)$   &-1640.1676 & 9.05 &-1640.1444 & 9.12 & 8.92 \\
 &$3d^{10}({}^1S)$  &-1640.2753 & 6.12 &-1640.2590 & 6.00 & 5.65 \\\hline\hline
Zn   &$3d^{10}4s^2({}^1S)$&-1779.4837 & 0.00 &-1779.4656 & 0.00 &      \\\hline
$\mbox{Zn}^+$&$3d^{10}4s({}^2S)$&-1779.1345 & 9.50 &-1779.1169 & 9.49 & 9.23 \\
&$3d^94s^2({}^2D)$ &-1778.8256 &17.91 &-1778.8048 &17.98 & 17.85 \\\hline\hline
\end{tabular}
\end{center}
\caption{ Results for B-LYP runs on the transition metals(continued).}
\label{tb:metals2}
\end{table}

\begin{table}[ht]
\begin{center}
\begin{tabular}{lrrrrrrr}\hline\hline
Ionization&HF&MP2&MP3&MP4&QCISD(T)&B-LYP\\\hline\hline
$d^ns^2\rightarrow d^ns$&1.43&0.32&0.31&0.10&0.09&0.10\\
$d^ns^2\rightarrow d^{n+1}$&0.75&0.52&0.43&0.38&0.14&0.22\\\hline
Excitations&&&&&&\\\hline
$d^ns^2\rightarrow d^{n+1}s$&0.77&0.49&0.53&0.41&0.12&0.51\\
$d^ns^2\rightarrow d^{n+2}$ &    &    &    &    &    &0.76\\\hline\hline
\end{tabular}
\end{center}
\caption{
Comparison of mean absolute deviations for the ionization and
excitation
energies of the first row transition metals.  The B-LYP results are
those of the current work, all others are taken from Raghavachari and
Trucks (RT)\protect\cite{raga-ions,raga-exciting}.  Errors are
relative to the experimental
values with relativistic corrections.}
\label{tb:compare-hfetc}
\end{table}
\begin{table}[ht]
\begin{center}
\begin{tabular}{lrrr}\hline\hline
Ionization&$\mbox{LDA}^a$&$\mbox{B-P}^b$&$\mbox{B-LYP}^c$\\\hline
$d^ns^2\rightarrow d^ns$&0.28&0.16&0.10\\
$d^ns^2\rightarrow d^{n+1}$&0.27&0.23&0.22\\\hline
\end{tabular}
\end{center}
\caption{
 Comparison of mean absolute deviations (eV) for the
ionization potentials of the first row transition metals.  ${}^a$
Slater Exchange with VWN correlation functional\protect\cite{ZiegLi} ${}^b$
Becke Exchange with Perdew Correlation Functional\protect\cite{ZiegLi}.
${}^c$ Present work.}
\label{tb:bp}
\end{table}
\begin{table}[ht]
\begin{center}
\begin{tabular}{lrrrr}\hline\hline
$d^9s^2\rightarrow d^{10}s$&   HF  &   B-null  & B-LYP  & B-LYP(HF)\\\hline
$\Delta E_{exch}$        &  -5.62 &  -5.69  &  -5.74  &  -6.99 \\
$\Delta E_{corr}$        &  -1.54 &   0.00  &  -0.16  &  -0.18 \\
$\Delta E_{coul}$        & 185.18 & 153.30  & 154.01  & 185.18 \\
$\Delta h_1$             &-179.88 &-149.50  &-150.16  &-179.88 \\
$E_{total}$              &        &  -1.89  &  -2.05  &  -1.86 \\\hline
$d^9s^2\rightarrow d^9s$ &       &           &        &        \\
$\Delta E_{exch}$        &   3.40 &   4.18  &   4.23  &   4.12 \\
$\Delta E_{corr}$        &   1.57 &   0.00  &   0.92  &   0.92 \\
$\Delta E_{coul}$        &-255.57 &-259.55  &-264.33  &-255.57 \\
$\Delta h_1$             & 259.52 & 263.60  & 268.23  & 259.52 \\
$E_{total}$              &        &   8.23  &   9.05  &   8.98 \\\hline
$d^9s^2\rightarrow d^{10}$&       &          &        &        \\
$\Delta E_{exch}$        & -2.51  &  -1.53  &  -1.49  &   2.90 \\
$\Delta E_{corr}$        & -0.45  &   0.00  &   0.40  &   0.31 \\
$\Delta E_{coul}$        &-60.38  & -95.15  & -98.52  & -60.38 \\
$\Delta h_1$             & 69.00  & 102.41  & 105.73  &  69.00 \\
$E_{total}$              &        &   5.73  &   6.12  &   6.02 \\\hline
\end{tabular}
\end{center}
\caption{
 Breakdown of energy contributions for B-LYP, B-null, HF and
B-LYP(HF) (B-LYP using Hartree-Fock density) calculations on Cu.  For
the HF case $E_{corr}$ is the empirical correlation, $E_{exp}-E_{HF}$}
\label{tb:pickapart}
\end{table}
\clearpage

\section{Figure Captions}

\nopagebreak

\end{document}